# Achieving Direct Electrochemical Oxidation of Carbon below 600ºC through a Novel Direct Carbon Fuel Cell

Wei Wu, Yunya Zhang, Yong Ding, Dong Ding, and Ting He
Energy & Environment Science and Technology
Idaho National Laboratory, Idaho Falls, ID 83415, USA

**Abstract**

Direct carbon fuel cells (DCFCs) are highly efficient power generators fueled by abundant and cheap solid carbons. However, the limited formation of triple phase boundaries (TPBs) within fuel electrodes inhibits their performance even at high temperatures due to the limitation of mass transfer. It also results in low direct-utilization of the fuel. To address the challenges of low carbon oxidation activity and low carbon utilization simultaneously, a highly efficient, 3-D solid-state architected anode has been developed to enhance the performance of DCFCs at intermediate temperatures (≤600°C). The cells with the 3-D textile anode, $Gd:CeO_2$-$Li/Na_2CO_3$ composite electrolyte, and $Sm_{0.5}Sr_{0.5}CoO_3$ (SSC) cathode have demonstrated excellent performance with maximum power densities of 143, 196, and 325 mW $cm^{-2}$ at 500, 550, and 600°C, respectively. At 500°C, the cells could be operated steadily with a rated power density of ~0.13 W $cm^{-2}$ at a constant current density of 0.15 A $cm^{-2}$ with a carbon utilization over 86%. The significant improvement of the cell performance at such temperatures attributes to the high synergistic conduction of the composite electrolyte and the superior 3-D anode structure which offers more paths for carbon catalytic oxidation. Our results indicate the feasibility of direct electrochemical oxidation of solid carbon at 500-600°C with a high carbon utilization, representing a promising strategy to develop 3-D architected electrodes for fuel cells and other electrochemical devices.

*Broader Context*

*Highly efficient and environmentally friendly utilization of solid carbons, the most abundant and cheapest energy source on Earth, is of great importance to meet the increasing global energy demand and environmental sustainability. DCFCs have gained much interest in recent decades due to the advantages of high efficiency, solid carbon fuel, and carbon dioxide capture ready. However, their performance is far beyond satisfaction due largely to insufficient contacts between carbon particles and conducting phases in the fuel electrode, which inhibits further development. Therefore, there exists an urgent and growing need for innovative approaches to develop DCFCs with higher power density and carbon utilization. For the first time, we report high performance DCFCs that are based on a 3-D architected textile anode and a highly conductive composite electrolyte operated below 600°C. The newly developed anode framework can potentially expedite the next-generation high performing fuel cells. Furthermore, the concept of the ceramic textile electrode framework can be readily applied to other energy systems, such as batteries, supercapacitors, and electrolyzers.*

## Introduction

Carbon, the main component of coal and biomass, is expected to continue dominating the power generation in developing countries because of its low price and ultra-high volumetric energy density.[1] Conventional power generation from carbon (coal) involves the combustion process and has a low energy-conversion efficiency due to the limitation of the Carnot cycle.[2] Solid oxide fuel cells, one of the most promising power generation technologies, has received considerable attention as clean alternatives for carbon, hydrogen, and hydrocarbon utilization, where chemical energy is converted into electric power electrochemically with high efficiency and low emissions.[3-5] One of special types of fuel cells is direct carbon fuel cells (DCFCs).

DCFCs are usually operated at temperature range of 700-900°C.[6, 7] High operating temperatures lead to fast reverse Boudouard reaction, which gasifies solid carbon into carbon monoxide in order to achieve the acceptable power output due to the enlargement of triple phase boundaries (TPBs). However, the gasification process decreases the energy conversion efficiency by using carbon monoxide as the fuel and thus makes the fuel cells not "true" DCFCs.[8] Moreover, issues associated with the high temperature operation, such as high degradation rate, sealing failure, utilization of expensive materials, slow response to rapid start-up, and poor thermal cycling remain challenging.

To reduce the operating temperature of DCFCs, it is commonly agreed that an advanced fast ionic conductor should be developed.[9-11] A promising approach is to apply solid composite electrolytes composed of molten hydroxides or carbonates and oxygen ion conductors at reduced temperatures.[10, 12-14] The conductivity of this kind of composite electrolyte, which is composed of a porous oxygen ion conducting phase such as gadolinium doped ceria (GDC) and a $CO_3^{2-}$ conducting molten carbonate phase, is much higher than that of the conventional solid electrolytes, such as yttria stabilized zirconia (YSZ) and doped ceria, at temperatures above 470°C.[15] It was also evidenced that direct electrochemical oxidation of solid carbon could occur at lower temperatures (<700°C) in DCFC systems when GDC was used as the electrolyte.[16] However, unlike hydrogen/hydrocarbons gas-fueled fuel cells, the primary challenge in achieving carbon direct oxidation at reduced temperature is effectively bringing solid carbon particles to the electrolyte/electrode interface and forming TPBs between solid fuel, anode, and electrolyte, where electrochemical oxidation can take place.[17, 18] Therefore, the design of fuel composition and anode structure is of ultimate importance for achieving high performing DCFCs at reduced temperatures.[7, 19]

Recently, approaches in developing novel carbon-based fuels with good mobility[20-22] and optimizing anode composition or microstructure[23-25] have been explored to improve electrode reactions. It was reported that dispersing solid carbon particulates in a molten carbonate can extend the anode active layer and improve its performance.[20, 26-28] When silver-based catalysts (Ag, $Ag_2O$, $Ag_2CO_3$) were added to the carbon-carbonate slurry, the performance of the DCFC could be further improved.[29] However, excess carbonate in cell anode may corrode the glass sealant, resulting in the loss of open circuit voltage (OCV).[30] To eliminate the negative impact of the corrosive carbonate, Yang et al. reported a carbon-air battery based on a solid oxide fuel cell integrated with a ceramic $CO_2$-permeable membrane that created a peak power density of 279.3 mW cm$^{-2}$ at 850°C.[31] Solid carbon was modified with a reverse Boudouard reaction catalyst to allow the gasification of carbon to carbon monoxide, which served as the fuel. Though the work provided a novel carbon-air battery concept for portable power applications, the carbon conversion rate was fairly low (14.63%) due to thermodynamic limitations. In addition to modifying the carbon fuel, optimizing anode microstructure is another effective way to improve the fuel oxidation kinetics in DCFCs. Li et al.[32] developed a porous Ni anode by filling nickel foam with carbon particles to improve the fuel-electrode physical contact. Although the cell performance was only 70 mW cm$^{-2}$, the idea of using ultra-porous anode to extend the length of TPB near electrode/electrolyte interface enhanced the cell power density.[5, 33] A 3-D fibrous porous electrode was applied by Liu's group[34], providing facile pathways for gas transport and efficient charge transfer and, thus, greatly enhancing the kinetics of electrode reactions. Additionally, highly active cathode materials with high oxygen reduction reaction (ORR) activity is crucial for operating DCFCs at lower temperatures. $Sm_{0.5}Sr_{0.5}CoO_3$ (SSC) has shown superior electrocatalytic activity and excellent stability at temperatures below 600°C.[35, 36] By applying SDC-embedded SSC composite fibers as the cathode, a maximum power density of 360 mW cm$^{-2}$ was achieved at 550°C in a solid oxide fuel cell (SOFC).[35] A $CO_2$-treated $SSC/Co_3O_4$ cathode was applied in a single-chamber fuel cell, with direct hydrocarbon as the fuel, operating at temperatures below 500°C.[37] A relatively low firing temperature of SSC, compared to the conventional cathodes such as $La_{0.8}Sr_{0.2}MnO_{3-\delta}$ and $La_{0.6}Sr_{0.4}Co_{0.2}Fe_{0.8}O_{3-\delta}$, could benefit the application of DCFCs fabricated at relatively low temperatures.[38]

In this work, we report a unique DCFC design consisting of a novel 3-D architected anode framework, a highly conductive GDC-carbonate composite electrolyte, and an SSC cathode, that has demonstrated excellent performance at temperatures below 600°C. At 500°C, the cell delivers a power density of 143 mW cm$^{-2}$ and good stability with Ar as the purge gas and 75%$CO_2$-25%$O_2$ as the oxidant. The mechanism of performance enhancement is also discussed.

## Experimental

**Fabrication of Carbonate-GDC Composite Electrolyte**

The fabrication of composite carbonate-GDC electrolyte was reported previously.[32] First of all, lithium–potassium carbonate, $Li_{0.67}K_{0.33}CO_3$, was formed by high-energy ball milling (SPEX SamplePrep LLC, NJ, USA) of $Li_2CO_3$ and $K_2CO_3$ in a mole ratio of 2:1 for 20 minutes, followed by calcination at 600°C for 2 hours. Secondly, composite electrolyte powder was obtained through mixing GDC powders and carbonate in a weight ratio of 3:7 and subsequent calcination at 700°C for 1 hour. After quenching, the composite powders were uniaxially pressed at 300 Mpa and fired at 750°C for 4 hours to form electrolyte pellets with a diameter of 10 mm.

**Fabrication of NiO-GDC 3-D Ceramic Textile Anode**

Figure 1 schematically illustrates the process for fabricating electrolyte-supported ceramic 3-D textile anode. The ceramic framework was fabricated through a template-derived firing procedure. NiO-GDC precursor solution was prepared by dissolving a stoichiometric amount of $Ni(NO_3)_2 \cdot 6H_2O$ (Sigma Aldrich), $Gd(NO_3)_3 \cdot 6H_2O$ (Sigma Aldrich), and $Ce(NO_3)_3 \cdot 6H_2O$ (Sigma Aldrich) in distilled water. A fabric textile (Telio, Montreal, CA) was immersed in the GDC precursor solution overnight, followed by firing at 750°C for 4 hours with a heating rate of 1°C·min$^{-1}$ to form NiO-GDC ceramic textile. Coupons with a diameter of 3/16 inch were then punched from the ceramic textile. The punched ceramic textile coupon was then bonded on the top of a prepared composite carbonate-GDC electrolyte using PVB/ethanol (10 wt%) solution with a loading of 30 mL cm$^{-2}$ and co-fired at 750°C for 2 hours to form a half cell with ceramic textile anode, as schematically shown in Figure 1(c). Graphite was used as carbon fuel in this research, with the weight ratio of carbon to $Li_2CO_3$–$K_2CO_3$ being 4:1. Diluted slurry containing a 0.01g mixture of carbon and carbonate particles was dropped on the top of 3-D anode frame followed by drying at room temperature overnight to let the carbon fuel go through the anode frame gaps and reach anode/electrolyte interface (Figure 1(d)).

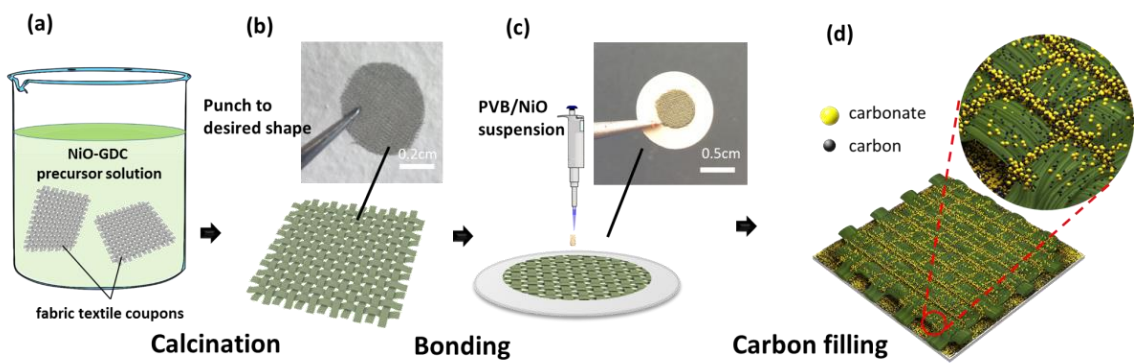

Figure 1. Process for fabricating 3-D ceramic textile anode framework: (a) fabric textile coupons soaked in NiO-GDC precursor solution; (b) NiO-GDC ceramic textile framework consisting of bundles of fibers; (c) bonding the 3-D anode on composite electrolyte surface by filling calcined frame with binder (PVB); and (d) schematic of 3-D anode with carbon-carbonate fuels. The particle size and carbonate/carbon ratio were adjusted in the figure to clearly demonstrate the fuel mixture and location.

**Cell Assembly and Testing**

The SSC powders were synthesized using a glycine-nitrate process.[39] The cathode (70wt% SSC and 30wt% GDC) was screen printed on the surface of the electrolyte pellet, followed by co-sintering at 750°C for 2 hours. The active area was 0.178 $cm^{-2}$. Button cells were sealed on an alumina tube using Aremco 552 sealant, with the anode side up. Silver mesh was used as current collectors with attached silver wires as leads. Ceramic cotton was stuffed in the tube near the carbon fuel to prevent the solid carbon from flowing away while reducing and purging with gas. Hydrogen was used during ramping up. After NiO was fully reduced to metallic Ni, Ar gas with a flow rate of 10 ml·$min^{-1}$ was swept in as purge gas. Oxygen and carbon dioxide (volume ratio of 1:3), with a total flow rate of 40 ml·$min^{-1}$, was fed as cathode gas. In electrolyte conductivity measurement, the composite electrolyte pellet was prepared by uniaxial pressing the premixed powders under a pressure of 280 MPa. Silver paste (Ag paste 9547, ESL ElectroScience Inc. USA) was printed on both sides of the pellet and fired at 700°C for 1 hour as current collectors. The measurements were carried out in the frequency range from 0.1 Hz to 1 MHz with a bias voltage of 10 mV. Cell I-V and I-P measurements, as well as electrochemical impedance spectroscopy, were recorded using a Solartron 1400 electrochemical working station when a stable OCV was observed. A schematic illustration of the cell testing configuration is shown in Figure S1.

**Characterization**

The phase purity of NiO-GDC textile was examined with a Rigaku SmartLab X-Ray Diffraction (XRD) in 15-90° angular range with 0.04° step size and a 1.6s resonance time.

The total conductivity of the carbonate-GDC electrolyte was measured in air in the temperatures range from 400–650°C using an electrochemical impedance spectroscopy (EIS) from Solartron (1400 Cell Test System). The textile anode microstructure, as well as cell cross-section, was characterized either via SEM (JEOL 6700F), equipped with back scattering electron (BSE) analyzer, or transmission electron microscope (TEM) equipped with energy dispersive X-ray spectroscopy (JEOL 4000 EX).

## Results and Discussion

**Characterization of 3-D Anode and Composite Electrolyte**

Fig. 2(a) shows a SEM image of a calcined 3-D anode framework. Bundles of NiO-GDC fibers were knit together forming a textile-like structure, ensuring not only the 3-D porosity but also a sufficient mechanical strength. An enlarged cross-sectional image is shown in Figure 2(b). The fibers are hollow with an average inner diameter of 1-2 μm, which allows molten carbonate assisted carbon particles to thoroughly infiltrate into the anode structure. The hollow fibers are formed by the outward diffusion of gases generated from the oxidization of the polymer additives as well as the decomposition of metal precursors during calcination. Similar hollow oxide fibers have also been reported via electrospinning technique.[40, 41] Micron-sized pores are evenly distributed throughout the wall of each fiber, as shown in Fig. 2(c), which further increase the specific surface area of hollow fibers and facilitate fast mass transfer. The detailed features of the hollow fibers are further revealed by scanning transmission electron microscopy (STEM). The zero energy loss energy-filtered image, Figure 2(d), and the electron energy-loss spectroscopy (EELS) mapping, Figure 2(e), from the spot shown in Fig. 2(c) show a relatively uniform distribution of Ni and Ce elements. Figure 2(f) is an EELS spectrum focused on the same position, in which the peaks corresponding to O, Ni, Ce, and Gd can be clearly identified indicating that the hollow fibers consist of uniformly distributed NiO and GDC particles. The phase purity of anode textile framework was examined with XRD. The standard phases of the NiO and GDC were confirmed (see Figure S2 in Supplementary Information), indicating that the fiber is well synthesized with no secondary phase.

The ionic conductivity of a dense GDC-carbonate composite electrolyte pellet was measured by EIS in a symmetric Ag/pellet/Ag configuration. The total resistance, including the bulk and grain boundary contributions, was calculated using low frequency intercept corresponding to the capacitive behavior of the Ag electrodes. The ionic conductivity was calculated using $\delta = L/(Z \times S)$, where Z is the impedance for the real axis in the Nyquist plot, L is the ceramic disk length, and S is the surface area, respectively. When the temperature is above 500°C, the total conductivity of the composite electrolyte is 0.047 S·cm$^{-1}$, indicating a factor of 15 and 235 times higher than GDC and YSZ, respectively. Our previous results suggest that the conductivity of the GDC-carbonate electrolyte highly depends on the volume fraction, especially at lower temperatures.[32] The ionic conductivity of the electrolyte against the inverse of temperature is plotted in the supplementary Figure S3.

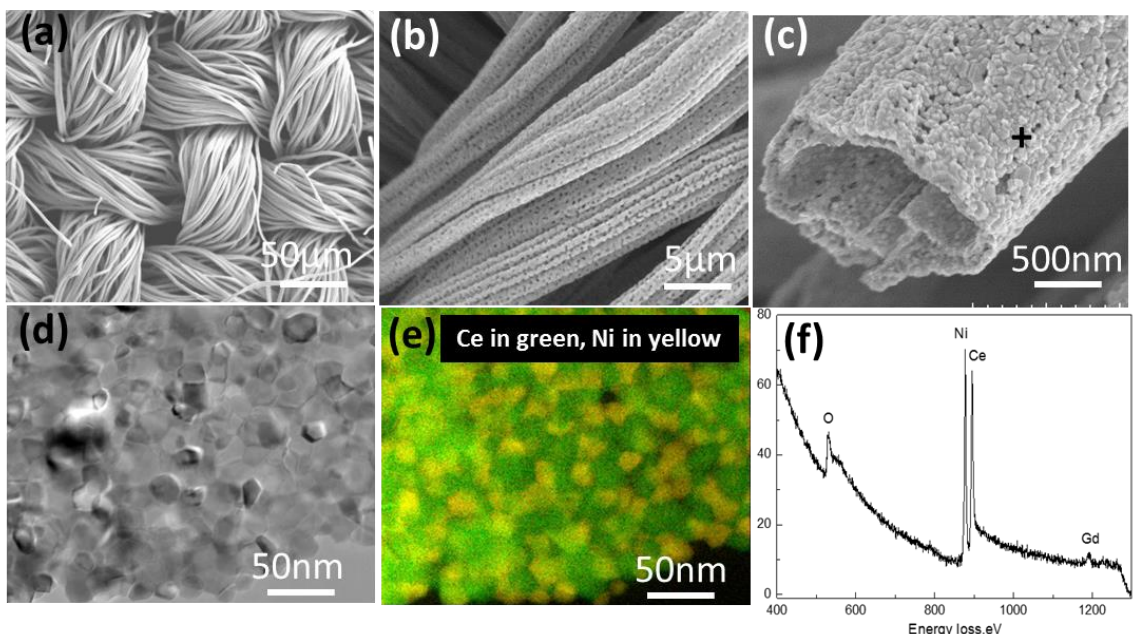

Figure 2. Three-D NiO-GDC anode framework: (a) SEM image of a 3-D NiO-GDC framework surface; (b) a closer view of the anode framework consisting of porous hollow fibers, showing open pores/channels for fast mass transfer; (c) an individual hollow fiber shown in (b); (d) HAADF STEM image; (e) EELS mapping; and (f) an EELS spectrum from the black cross spot marked on the wall of fiber in (c).

**Electrochemical Performance**

Figure 3(a) shows typical I-V polarization curves of the DCFCs operated between 500 and 600°C with the 3-D anode frameworks. The OCVs stay approximately at 1.19 V, 1.18 V, and 1.16 V for operations at 500, 550, and 600°C, respectively. Theoretically, the OCV of a DCFC with oxygen as oxidant is independent to temperature because carbon is always in its elementary state,[8] and an equilibrium cell potential of 1.02 V should be expected in accordance with complete carbon oxidation reaction:[17]

$$C + O_2 \rightarrow CO_2 \qquad [1]$$

However, this research did show temperature dependence of cell potentials, and the OCVs are much higher than the equilibrium potential of 1.02V. Same phenomena were also reported by others on carbon fuel cell where carbonates were involved in carbon conversion.[42, 43] The molten carbonates in the electrolyte at the operating temperature was proposed to keep the activity of electrochemical products low and contribute to high OCV.[44] Furthermore, the molten state of carbonates in the electrolyte and anode will help eliminate potential issues of gas leakage through electrolyte, resulting in high cell open circus voltage. We also found that the purge gas argon enhances open circuit voltage by ~0.15 V due to the reduced partial pressure of $CO_2$, which is consistent with the literature that reported an enhancement of 0.1-0.3 V to the equilibrium cell potential.[45] The maximum power densities reached 325 mW cm$^{-2}$ and 196 mW cm$^{-2}$ at 600°C and 550°C, respectively. When the temperature was further reduced to 500°C, which is close to the melting point of $Li_2CO_3$–$K_2CO_3$ carbonate, a maximum power density of 143 mW cm$^{-2}$ could still be achieved. This power density is 2.8 times the performance of GDC-based DCFC at 700°C, which was reported to be 50 mW cm$^{-2}$.[16] It is also higher than the performance of the electrolyte-supported cell operated at 650°C reported by Xu. et al.[46], which has a peak power density of 113.1 mW cm$^{-2}$. Figure 3(b) displays the impedance spectra of a cell under OCV condition at different temperatures. The ohmic resistances of the cell, $R_s$, corresponding to the high frequency intercepts of the impedance spectra with the real axis in the Nyquist plots, are 0.28, 0.20, and 0.15 Ω cm$^2$ at 500, 550, and 600°C, respectively. The intercepts in the low-frequency region are total resistances, including $R_s$ and the polarization resistance $R_p$. Thus the $R_p$ can be obtained by subtracting $R_s$ from the total resistance. The calculated $R_p$ for the cell before stability test are 3.27, 2.40, and 1.55 Ω cm$^2$ at 500, 550, and 600°C, respectively. Although both ohmic and polarization resistances demonstrated a decrease with the increase of temperature, as shown in Figure 3(c), their contributions to total resistance change differently. The ratio of ohmic resistance to total resistance decreases with the drop of temperature, while that of polarization resistance increases (Table 1). Thus, reducing polarization resistance is important in further improving cell performance at lower temperatures. The cell could be operated at a constant current density of 0.15 A cm$^{-2}$ for approximately 123 min followed with a sharp voltage drop (Figure 3(c)) due to limited fuel amount. Approximately 197 Coulombs of charge were released during the stable operation, which is equivalent to the amount of electricity generated by 0.0061 g carbon through electrochemical reaction. Considering there was 0.008 g carbon (80wt% of 0.01 g carbon-carbonate composite fuel) initially loaded in the 3-D anode prior to the test, approximately 76.2% of the total carbon fuel was converted to electricity. The fuel utilization is approximately 4.6 times that of the fuel cell using converted CO gas as fuel.[31] A comparison of power densities (Figure 4) clearly depicts how this work stands out from those reported in literature. As shown in the purple area in Figure 4, most DCFCs operated at the temperature range between 650 and 900°C with limited power densities. In 2011, Jayakumar et al., reported a DCFC using molten Sb-$Sb_2O_3$ mixture as anode with a maximum power density of 350 mW cm$^{-2}$ at 700 °C.[25] However, the OCV of that DCFC was low due to the intrinsic low Nernst potential for the Sb-$Sb_2O_3$ equilibrium. One year later, Jiang et al., reported a carbon fuel cell that exhibited a maximum power density of 390 mW cm$^{-2}$ at 750 °C with a lanthanum doped strontium manganite (LSM) cathode.[28] As the author mentioned, the significant improvement of power density is thought to result from the following factors: lower ohmic resistance and polarization resistance, less carbonate blocking gas/liquid/solid interfaces, faster gas diffusion and transportation, highly catalytic cathode materials, flowing gas at the cathode, etc. Considering the reverse Boudouard reaction at the operating temperature (750°C), the enhanced performance was due to the higher amount of CO formation and its electrochemical oxidation at the anode. Our research toward low temperature DCFCs provides remarkable evidence for direct electrochemical oxidation of carbon and established a benchmark below 600°C, opening a unique approach in developing high performance DCFCs at reduced temperatures.

The high performance at lower operating temperatures results from the unique 3-D anode framework and superior ionic conductivity of the composite electrolyte. The 3-D anode textile framework offers more pathways for carbon particles to reach the active reaction zone. The formation of triple phase boundaries through sufficient contacts between carbon fuel and electrolyte is commonly agreed to as the best approach to improve fuel oxidation and cell performance.[47]

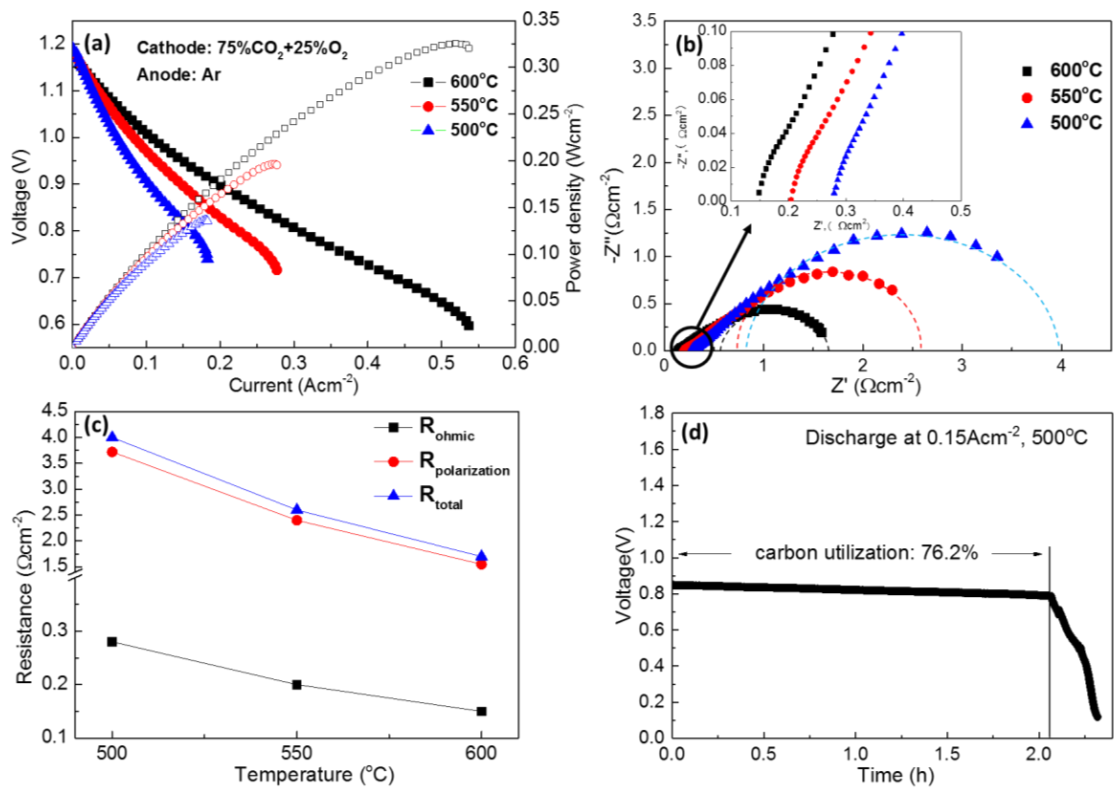

Figure 3. (a) Polarization and power density curves of DCFCs operated between 500 and 600°C; (b) The corresponding impedance spectra; (c) Ohmic resistance ($R_s$), polarization resistance ($R_p$) and total resistance ($R_{total}$) as a function of temperature; and (d) Long-term stability of a cell operated at 500°C at a constant current density of 0.15A cm$^{-2}$. Argon was used as purge gas.

Table 1. The variation of ohmic and polarization resistances versus temperature.

|  | $R_s$/Ωcm$^{-2}$ | $R_p$/Ωcm$^{-2}$ | $R_{total}$/Ωcm$^{-2}$ | $R_s$/$R_{total}$ | $R_p$/$R_{total}$ |
|---|---|---|---|---|---|
| 500°C | 0.28 | 3.72 | 4.0 | 7.0% | 93.0% |
| 550°C | 0.2 | 2.4 | 2.6 | 7.7% | 92.3% |
| 600°C | 0.15 | 1.55 | 1.7 | 8.8% | 91.2% |

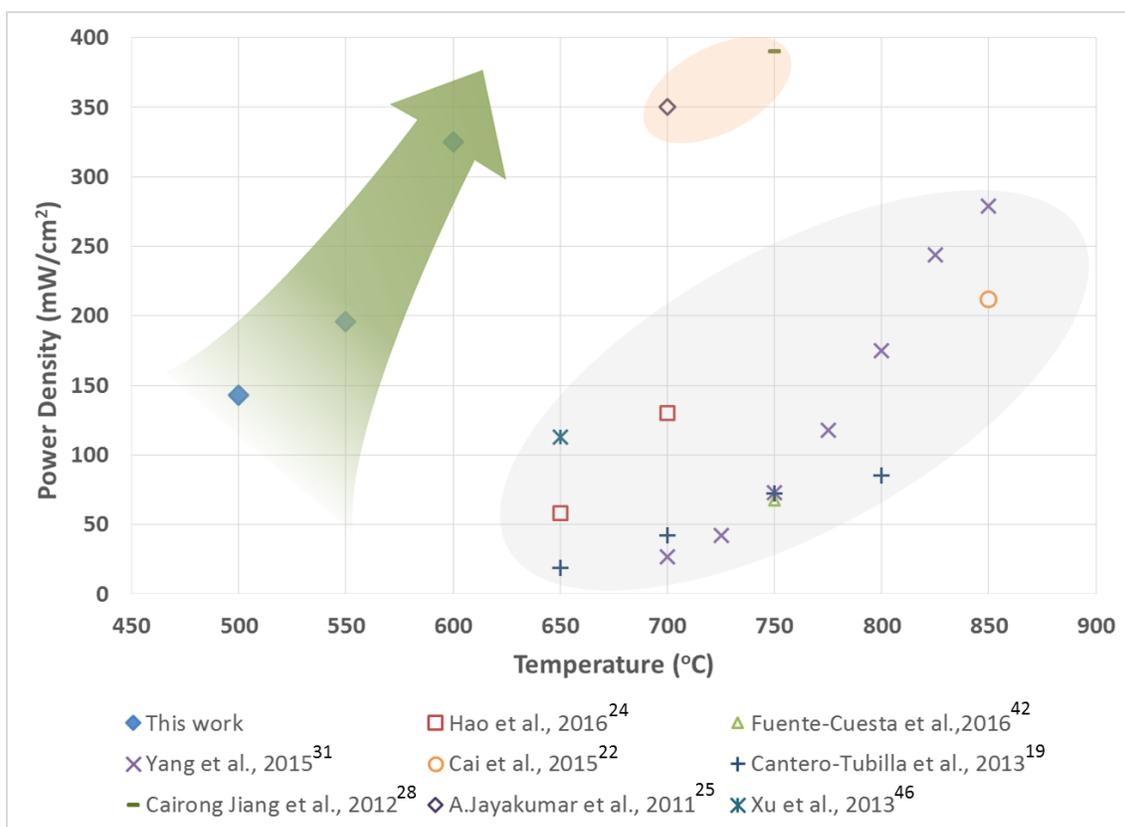

Figure 4. Performance comparison of this work with literatures.

**Mass Transfer and Reaction Mechanisms**

Figure 5 (a) and (b) show SEM images of the 3-D anode framework before loading carbon fuel and after short electrochemical testing. The bundle of hollow fibers were bonded very well with the GDC-carbonate electrolyte, establishing excellent contacts between electrode and electrolyte, a critical step for enhancing charge transfer. After carbon fuel was loaded, the carbon particles went through the gaps between fibers and reached the electrolyte/electrode interface. The addition of the carbonate in carbon fuel actually helped extend the electrolyte region. During the operation, the added molten carbonates carried carbon particles further into the hollow fibers through the pinholes on the fiber wall, as shown in Figure 5(b). The issue of carbon distribution within the anode was well addressed by applying this unique 3-D architected hollow fiber anode framework. Figure 5(c) illustrates the electrochemical processes near the electrolyte-anode interface. There have been a number of studies investigating the electrochemical oxidation of carbon in carbonate melts.[42, 48-51] However, the mechanism is not yet fully understood. Various aspects such as the role of reverse Boudouard reaction, as well as the decomposition and carbothermic reaction of carbonates, are still under investigation.[8] In the present cell configuration, both $CO_3^{2-}$ and $O^{2-}$ ions are charge species in the composite electrolyte. Oxygen ions are conducted along the GDC bulk phase and grain boundary whereas carbonate ions transfer through the molten carbonate phase.[52] As the operating temperature is no higher than 600°C, the influence of reverse Boudouard reaction can be neglected because the reverse Boudouard reaction is thermodynamically unfavorable below 700°C.[44]

Therefore, in this research, the anode reaction mechanism can be simplified as direct carbon electrochemical oxidation. At the interface between electrolyte and anode, the carbon particles directly contact the GDC phase and react with $O^{2-}$ to produce carbon dioxide and release electrons, whereas those contacting the carbonate phase react with $CO_3^{2-}$ ions and produce carbon dioxide and electrons, simultaneously. As presented in Figure 2, the 3-D anode hollow fibers consist of GDC and Ni phases, which provide abundant sites for carbon particles to react with $O^{2-}$ ions. The carbonate addition in the carbon fuel will provide reaction pathways for carbon to react with $CO_3^{2-}$. In cathode, the oxygen can be directly reduced into oxygen ions and also react with $CO_2$ to generate carbonate ions. The anode and cathode reactions can be expressed as the following:

Cathode side reaction:
$$O_2 + 4e^- \rightarrow 2O^{2-} \quad [2]$$
$$O_2 + 2CO_2 + 4e^- \rightarrow 2CO_3^{2-} \quad [3]$$

Anode side reaction:
$$C + 2O^{2-} \rightarrow CO_2 + 4e^- \quad [4]$$
$$C + 2CO_3^{2-} \rightarrow 3CO_2 + 4e^- \quad [5]$$

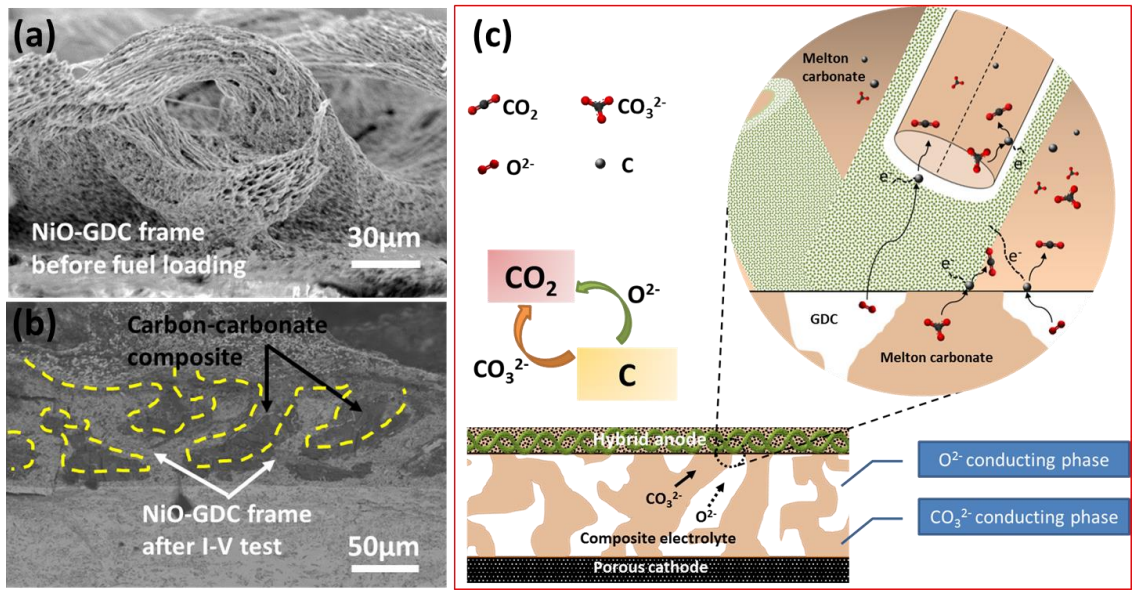

Figure 5. SEM images of anode framework (a) before loading carbon and (b) after short electrochemical testing. (c) Electrochemical process at electrolyte-electrode interface.

In the conventional carbonate-based DCFCs, where the carbonates were restricted in electrolyte layer only, the power output is highly limited by the mass transfer within the anode, especially near the molten point of carbonates.[8] The lack of TPBs between carbon and ion conducting phases results in a very low carbon utilization.[7] Our research approach offers a feasibility of directly electrochemical oxidation of solid carbon in a DCFC at 500-600°C with high carbon utilization. Further modification of cathode will help improve the cell performance and the research is ongoing. Noted that the fuel is "fluid" like at operating temperature at the present work, the continuous feed-in of the fuel is feasible so that a "true fuel cell" rather than a "battery" can be realized.[53][31]

## Conclusions

High performance DCFCs operated at temperatures ≤ 600°C have been demonstrated by applying NiO-GDC 3-D ceramic textile hollow framework as anode, GDC-Li/Na$_2$CO$_3$ composite as electrolyte, and SSC as cathode with graphitic carbon as fuel. Power densities of 325 mW cm$^{-2}$, 196 mW cm$^{-2}$, and 143 mW cm$^{-2}$ were obtained at 600, 550, and 500°C, respectively. The cell showed reasonable output stability with a carbon utilization over 86%. The porous 3-D anode structure offers a great ability to "absorb" carbon particles for direct electrochemical oxidation. Our results present a good prospective in promoting the direct carbon oxidation through DCFCs at reduced temperatures.

## Acknowledgements

The authors gratefully acknowledge the Idaho National Laboratory Directed Research and Development Program under DOE Idaho Operations Office Contract DE-AC07-05ID14517 for the support of this work.

# A High-Performing Direct Carbon Fuel Cell with 3-D Architectured Anode Operated below 600°C

Wei Wu, Yunya Zhang, Dong Ding, and Ting He
Energy and Environment Science and Technology
Idaho National Laboratory, Idaho Falls, ID 83415

## Supporting information

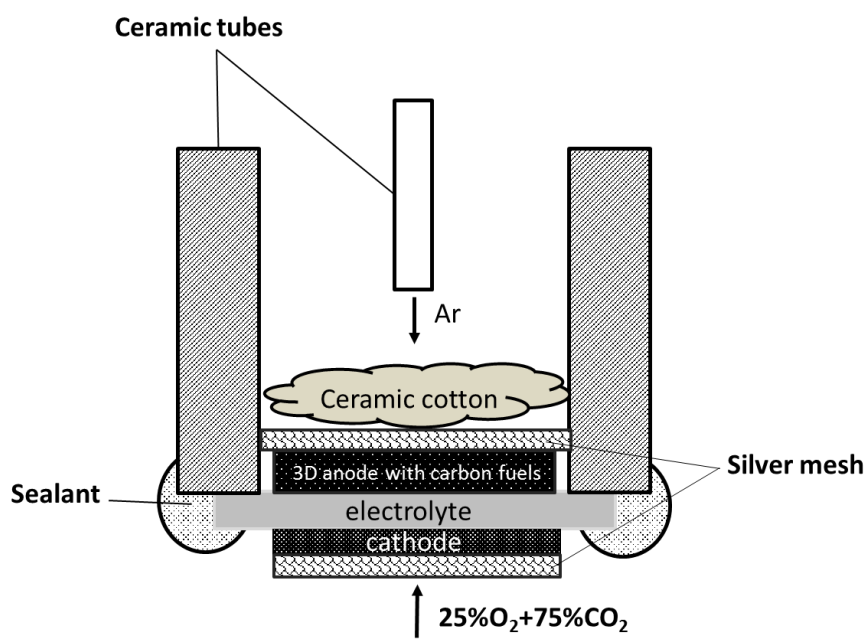

Figure S1. Schematic illustration of DCFC testing set up in this research.

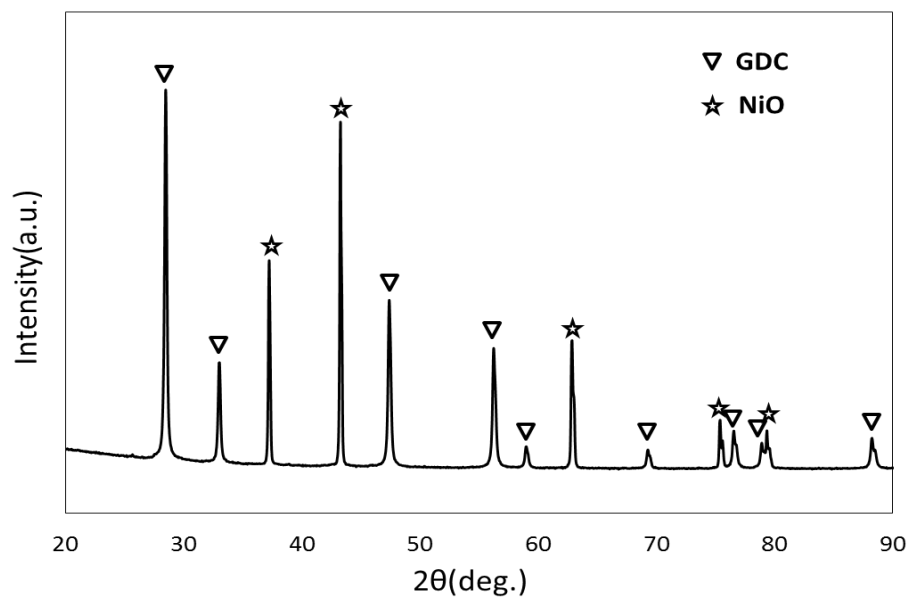

Figure S2. XRD pattern of a NiO-GDC textile framework.

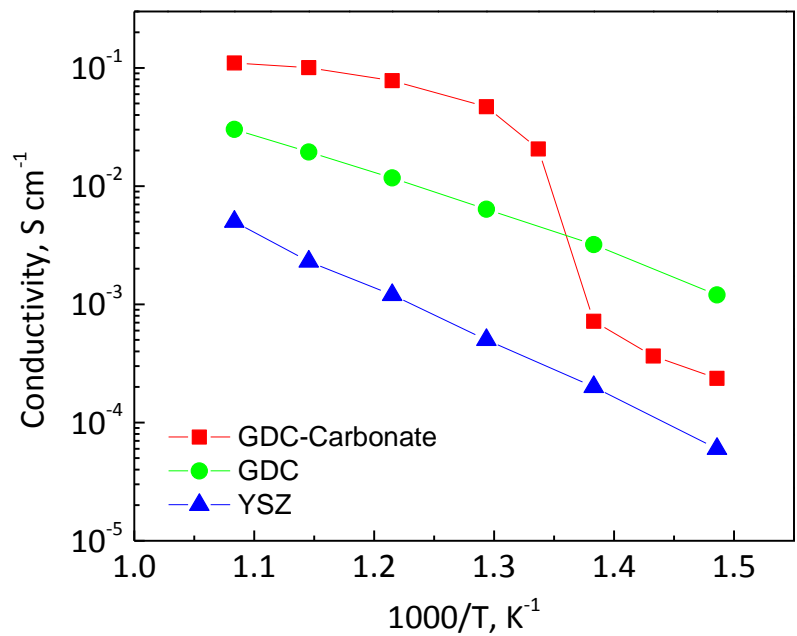

Figure S3. Comparison of ionic conductivities of GDC-carbonate, GDC and YSZ in air.

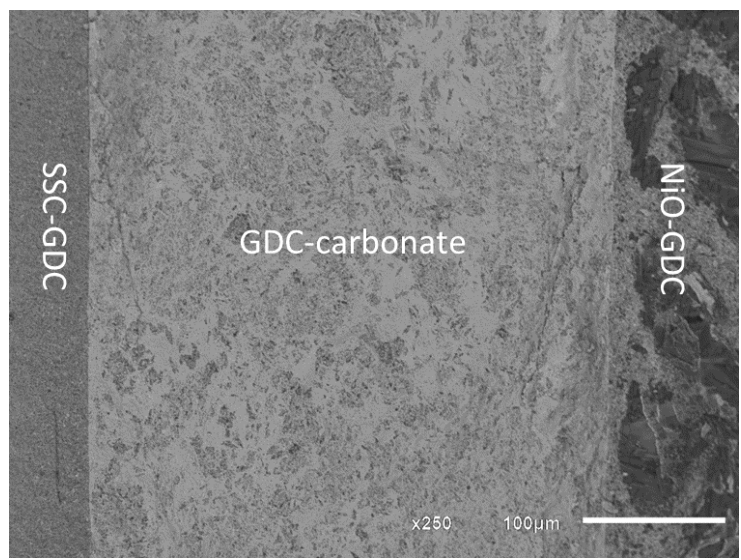

Figure S4. SEM cross-sectional image of a direct carbon fuel cell post performance testing.

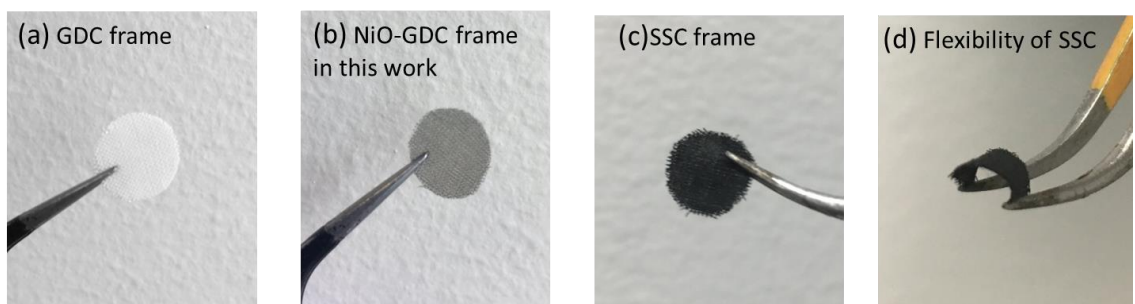

Figure S5. Capability in preparing various ceramic based 3D framework for electrochemical applications: (a) GDC textile framework; (b) NiO-GDC framework used in this work; c)$Sm_{0.5}Sr_{0.5}CoO_3$; and(c) good flexibility of SSC framework.